\documentclass[10pt,twoside,BCOR7mm,DIV15,headinclude,footexclude,
               cleardoubleempty,idxtotoc]
{scrartcl}

\usepackage{natbib}
\usepackage[font=small,labelfont=bf]{caption}
\usepackage[english]{babel}
\usepackage{graphicx}
\usepackage{hyperref}
\usepackage{scrpage2}
\usepackage{ifthen}
\usepackage{booktabs}
\usepackage{amsmath}
\usepackage{amssymb}
\usepackage{multicol}
\usepackage{float}
\usepackage{hyperref}

\hypersetup{breaklinks=true
,colorlinks=true,linkcolor=black,urlcolor=blue
,citecolor=black}

\addto\captionsenglish{%
}

\makeatletter
\renewcommand{\@biblabel}[1]{}
\renewcommand{\@cite}[2]{%
{#1\ifthenelse{\boolean{@tempswa}}{,#2}{}}}
\makeatother
\ofoot{\thepage}
\ifoot{}

\setcapindent{0em}
\setheadsepline{1pt}

\setkomafont{pagehead}{\normalfont\sffamily}
\setkomafont{pagenumber}{\normalfont\rmfamily}

\makeatletter
\newcommand{\listofcontributions}{\@starttoc{con}}

\newcommand{\l@contribution} {\@dottedtocline{1}{1.5em}{2.3em}}
\makeatother

\newenvironment{contribution}{
\setcounter{section}{0}
\setcounter{figure}{0}
\setcounter{table}{0}
}{
\newpage
\lehead{}
\rohead{}
}

\begin{document}

\setlength{\baselineskip}{2.5ex}

\begin{contribution}

\newcommand{\email}{michael.f.corcoran@nasa.gov}
\newcommand{\axaf}{{\it CHANDRA}}
\newcommand{\chandra}{{\it CHANDRA}}
\newcommand{\rxte}{{\it RXTE}}
\newcommand{\asca}{{\it ASCA}}
\newcommand{\rosat}{{\it ROSAT}}
\newcommand{\einstein}{{\it EINSTEIN}}
\newcommand{\ginga}{{\it GINGA}}
\newcommand{\bbxrt}{{\it BBXRT}}
\newcommand{\suzaku}{{\it Suzaku}}
\newcommand{\xmm}{{\it XMM}}
\newcommand{\sax}{{\it BeppoSAX}}
\newcommand{\ec}{$\eta$~Car}
\newcommand{\glast}{{\it GLAST}}
\newcommand{\swift}{{\it Swift}}
\newcommand{\integral}{\textit{INTEGRAL}}
\newcommand{\nustar}{\textit{NuSTAR}}
\newcommand{\ms}{$M_{\odot}$}
\newcommand{\rs}{$R_{\odot}$}
\newcommand{\ls}{$L_{\odot}$}
\newcommand{\kms}{km~s$^{-1}$}
\newcommand{\fluxcgs}{ergs~s$^{-1}$~cm$^{-2}$}
\newcommand{\lumcgs}{ergs~s$^{-1}$}
\newcommand{\hst}{{\it HST}}

\lehead{M. F.\ Corcoran, K.\ Hamaguchi \& J. K.\ Liburd, et al.}

\rohead{The X-ray Lightcurve of $\eta$~Car, 1996--2014}

\begin{center}
{\LARGE \bf The X-ray Lightcurve of the Supermassive star $\eta$~Carinae, 1996--2014}\\
\medskip

{\it\bf M. F.\ Corcoran$^{1}$, 
         K.\ Hamaguchi${^2}$ ,
         J. K.\ Liburd$^{3}$, 
         D.\ Morris$^{3}$
         T. R.\ Gull$^{4}$, 
         T. I. Madura$^{1}$, 
         M.\ Teodoro$^{4}$, 
         A. F. J.\ Moffat$^{5}$, 
         N. D.\ Richardson$^{5}$, 
         C.~M.~P.~Russell$^{6}$, 
         A.~M.~T.~Pollock$^{7}$, 
         S.~P.~Owocki$^{8}$}


{\it              $^1$NASA \& USRA; Code 662, GSFC, Greenbelt MD USA                                }    \\
{\it              $^2$NASA \& UMBC; Code 662, GSFC, Greenbelt MD USA                                }    \\
{\it			  $^{3}$University of the Virgin Islands, St. Thomas, Virgin Islands, USA   }    \\
{\it			  $^{4}$ NASA \& Western Michigan University, Kalamazoo, MI, USA                            }    \\
{\it			  $^{5}$Universit\'e de Montr\'eal, Montreal, QC, Canada                        }    \\
{\it		      $^{6}$NASA \& ORAU; Code 667, GSFC, Greenbelt, MD, USA                        }    \\
{\it			  $^{7}$ESA, Vilspa, Spain.                                                 }    \\
{\it			  $^{8}$University of Delaware, Newark, DE, USA                             }

\begin{abstract}
Eta Carinae (\ec) is the nearest example of a supermassive, superluminous, unstable star. Mass loss from the system is critical in shaping its circumstellar medium and in determining its ultimate fate. Eta Car currently loses mass via a dense, slow stellar wind and possesses one of the largest mass loss rates known. It is prone to episodes of extreme mass ejection via eruptions from some as-yet unspecified cause; the best examples of this are the large-scale eruptions which occurred in 19th century. Eta Car is a colliding wind binary in which strong variations in X-ray emission and in other wavebands are driven by the violent collision of the wind of \ec-A and the fast, less dense wind of an otherwise hidden companion star. X-ray variations are the simplest diagnostic we have to study the wind-wind collision and allow us to measure the state of the stellar mass loss from both stars. We present the X-ray lightcurve over the last 20 years from \textit{ROSAT} observations and monitoring  with  the \textit{Rossi X-ray Timing Explorer} and the X-ray Telescope on the \textit{Swift} satellite. We compare and contrast the behavior of the X-ray emission from the system over that timespan, including surprising variations during the 2014 X-ray minimum.

\end{abstract}
\end{center}

\begin{multicols}{2}

\section{Unraveling the Nature of $\eta$ Car}

Eta Car is one of the most fascinating stars in the Milky Way, and the most luminous and massive object inside 10,000 lightyears. A series of eruptions and mass ejections in 1838--1843 \citep{1838MNRAS...4..121H,1843MNRAS...6....9M} having a kinetic energy rivalling a supernova brought the star to prominence, though its nature was a matter of contentious debate for more than a century.  \cite{1996ApJ...460L..49D} provided a crucial clue in the interpretation of the system by recognizing a 5.52-year period in the strength of HeI~10830\AA\ emission features correlated with NIR variations and which he ascribed to ``S-Doradus'' variations seen in other luminous, blue, and variable stars.  X-ray variability was first reported by \cite{Corcoran:1995fk}, and comparisons of the X-ray spectrum of \ec\ to spectra of  colliding wind binary systems, notably HD 193793=WR 140, were made as early as 1996 \citep[for example][]{1998ApJ...494..381C}.  Figure~\ref{mfcorcoran-fig:specvar} shows the X-ray variation observed by the PSPC, compared to high-resolution \chandra\ spectra obtained at key times.  Radial velocity variations in the Pa-$\gamma$ line suggested binary orbital motion \citep{1997NewA....2..107D}, though the true nature of the star as a colliding wind binary system was only confirmed by variations in the X-ray lightcurve \citep{1999ApJ...524..983I}.

\begin{figure}[H]
\begin{center}
\includegraphics[width=\columnwidth]{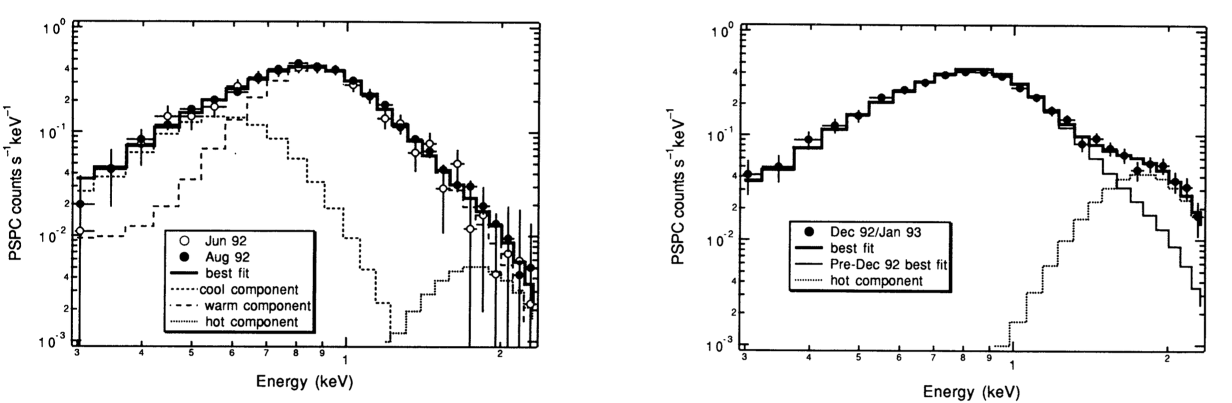}
\includegraphics[width=\columnwidth]{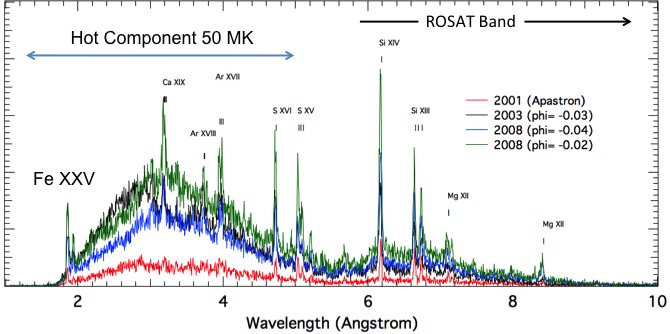}
\caption{Top: X-ray spectral variation from 1992 as seen by the \rosat\ PSPC \citep{Corcoran:1995fk}. Bottom: X-ray spectral variations from high spatial- and spectral- resolution observations with the \chandra\ High Energy Transmission Grating Spectrometer.  Strong resolved line emission shows the thermal nature of the source and fixes the temperature at about 4.5keV, corresponding to a pre-shock stellar wind velocity of about 3000~\kms.  The \rosat\ band, corresponding to 0.5-2.0 keV (6.0-24\AA) is also shown for comparison.
\label{mfcorcoran-fig:specvar}}
\end{center}
\end{figure}

\begin{figure*}[!t]
\begin{center}
\includegraphics
  [width=\textwidth]{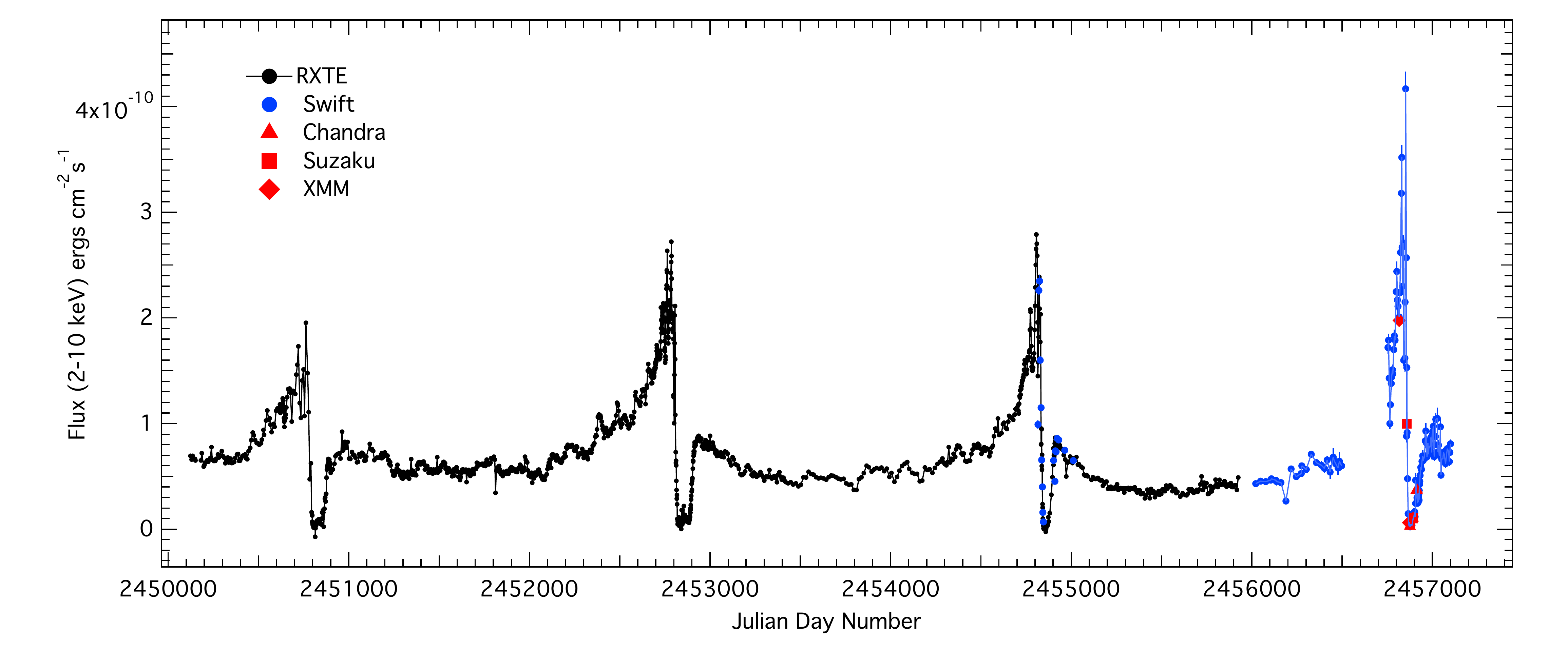}
\caption{Flux Variations of \ec, 1996-2014.
\label{mfcorcoran-fig:lc}}
\end{center}
\end{figure*}

\section{The X-ray Lightcurve Since 1996}

%

The recognition of a strict period by Damineli in the 1990's, along with the (serendipitous) discovery of \ec's X-ray variability by \rosat, and the suspicion that the broad-band variations could be driven by binary interactions,  made monitoring observations at X-ray energies essential.  
Fortuitously, this period also co-incided with the 
launch of the \textit{Rossi X-ray Timing Explorer} \citep[\rxte;][]{rxte} in December, 1995. 
%
%
\rxte, and its workhorse instrument, the Proportional Counter Array (PCA), provided for the first time the capability of monitoring temporal changes of a wide variety of cosmic X-ray sources, a crucial capability needed to show the full extent and nature of the variation of \ec's X-rays.  While \rxte\ was not very sensitive to the soft X-ray emission from most stellar sources, the PCA did have sufficient sensitivity to measure the relatively hard X-ray emission from long-period colliding wind binaries. 
The first observation of \ec\ by \rxte\ was on 09-February-1996, barely two months after \rxte's launch, while the final observation of \ec\ by \rxte\ occurred on 28-Dec-2011, just one week before the lamented end of the \rxte\ mission on 04-January-2012. 

After the demise of the \rxte, monitoring observations were obtained by the X-Ray Telescope (XRT) on NASA's \swift\ high-energy space observatory \citep{Burrows:2005qy}. The primary mission of \swift\ is to observe and localize Gamma-ray bursts, but, when not observing bursts, its XRT may be used to obtain spatially-resolved X-ray spectra of other Galactic and extra-galactic objects.  Observations of \ec\ with the XRT actually started in 2009, prior to the end of the \rxte\ mission, which provided good cross-calibration between the PCA and the XRT.  After a brief hiatus, XRT monitoring observations of \ec\ resumed on 03-April-2012 until 23-July-2013, then again from 05-April-2014 to 18-March-2015. 
%
Figure \ref{mfcorcoran-fig:lc} shows the X-ray flux vs. time as obtained by the \rxte-PCA and \swift-XRT, along with ``snapshot'' fluxes obtained with \chandra, \xmm, and \suzaku.  


\section{Understanding the X-ray Variations and \ec}

The variation of the X-ray spectrum as measured (crudely) by the PCA and (more fully) by \swift, and \chandra, \xmm, \suzaku\ (and now \nustar), provides a rich set of data that can be used to constrain the mass-loss rate from the primary star (\ec-A), the mass loss rate of the companion star (\ec-B), and the wind speed of the companion star.  Interpretations of the X-ray variations \citep{1999ApJ...524..983I, 2001ApJ...547.1034C, 2008MNRAS.388L..39O, 2009MNRAS.394.1758P,2009MNRAS.397.1426K,Russell:2011fk, Madura:2013fj} 
have led to a fairly simple concept of the system, in which \ec-A is an extremely massive, superluminous primary star with a strong ($\dot M\approx10^{-4}$\ms), slow ($V_{\infty,A}\approx 500$~\kms) wind, that is orbited by a hotter, less massive star with a weaker ($\dot M\approx10^{-5}$\ms), faster ($V_{\infty,B}\approx 3000$~\kms) wind, in an extremely eccentric ($e\sim0.9$), long period ($P=2023$ day) orbit. 
Near apastron, the companion's wind carves a large cavity in the wind of \ec, allowing UV radiation from the companion star and from the inner wind of \ec-A to escape in the direction of the observer at earth.  Near periastron, there's a fast transition from X-ray maximum to minimum (and from a high to low nebular ionization state) as the wind-wind cavity moves behind \ec-A and is swallowed up by the massive primary wind.  The cavity produced by \ec-B in the massive wind is the root cause of all the periodic spectral variations that have been observed from \ec\ over almost the entire electromagnetic spectrum.  

\section{Puzzles}

Although the colliding wind binary model provides a good foundation on which to interpret the X-ray variations of \ec\ (and the variability at other wavelengths) there are a number of important mysteries still to be solved.  The primary puzzle is the identity of the companion star, whose direct detection has remained frustratingly difficult.  The X-ray spectrum, which provides the most direct information about the companion through characterization of its wind properties, requires that \ec-B have an unusually high wind terminal velocity, and exceptionally high mass-loss rate.  Naively, a wind velocity of $V_{\infty}\approx 3000$~\kms and a mass loss rate of $\dot M\approx10^{-5}$\ms\ would indicate a very bright O star or perhaps an Of/WN star; if so this would make \ec-B one of the most luminous stars in the Carina Nebula (and easily detected if not for the glare from \ec-A).  The X-ray spectral variations themselves are not simple, and show secular (non-phase-locked) changes, for example the X-ray ``flares'' discussed by \cite{1997Natur.390..587C} and \cite{2009ApJ...707..693M}, and which seem unique to \ec.   The duration of the X-ray minimum itself has been seen to vary 
\citep{2010ApJ...725.1528C}.
The most recent periastron passage, which occurred in 2014, showed a longer recovery somewhat intermediate between the quick recovery of 2009 and the 1998 and 2003 X-ray minima.   The long period, high-eccentricy orbit is unusual,  though not unprecedented, since we know of one other established colliding wind binary, WR 140, which has a (suspiciously?) similar orbit.  The orbit as we understand it today must provide some clue as to the formation of the binary, perhaps by stellar capture or perhaps by some energetic event (the Great Eruption?) which drove the system almost to the breaking point.  

The relation between the binary system as we see it today, and the ``Great Eruption'' of 1838-1843, and the ``Lesser Eruption'' of 1890, is 
a tantalizing puzzle.  The series of eruptions in the 5 year period 1838-1843, and the 1890 eruption may be associated with periastron passages \citep{1997NewA....2..107D,Smith:2011uq}; if so, this suggests that binary interactions when the two stars are close may have helped modulate (or even trigger?) the eruptive event. It may be that the system was driven to its current orbital configuration, i.e., large semi-major axis and high eccentricity, by the eruptive mass loss of the 19th century.  The dynamics of the system before, during, and after the Great Eruption are interesting, and bear further scrutiny.  

\section{Summary}

X-ray flux monitoring with \rxte\ and \swift\ has provided the key to establishing \ec\ as a colliding wind binary system, since the X-ray spectral variations of \ec\ are analogous to other well know colliding wind binary systems and are thus straightforwardly interpreted. Monitoring observations like these provide important diagnostics of the orbital parameters of and the state of mass loss (and possibly mass transfer) from these massive stars. 

\bibliographystyle{aa} 
\bibliography{myarticle}

\end{multicols}

\end{contribution}


\end{document}